\documentclass[preprintnumbers,prd,onecolumn,floatfix,superscriptaddress, nofootinbib]{revtex4-2}

\usepackage{setspace}
\usepackage{graphicx}
\usepackage{subcaption}
\usepackage{epsfig}
\usepackage{bm}
\usepackage{amssymb}
\usepackage{float}
\usepackage{amsmath}
\usepackage{dcolumn}
\usepackage[usenames,dvipsnames]{color}
\usepackage{enumitem}
\usepackage[mathscr]{eucal}
\usepackage{hyperref}
\def\doi{http://doi.org}

\newcommand{\be}{\begin{equation}}
\newcommand{\ee}{\end{equation}}
\newcommand{\beano}{\begin{eqnarray*}}
\newcommand{\eeano}{\end{eqnarray*}}
\newcommand{\ba}{\begin{eqnarray}}
\newcommand{\ea}{\end{eqnarray}}

\def\chariteratehelpA#1 #2\relax{%
  \chariteratehelpB#1\relax\relax%
  \ifx\relax#2\else\rlap{\charop{~}}\ \chariteratehelpA#2\relax\fi
}
\def\chariteratehelpB#1#2\relax{%
  \charop{#1}%
  \ifx\relax#2\else
    \chariteratehelpB#2\relax%
  \fi
}
\def\charop#1{\def\stacktype{L}\def\useanchorwidth{T}%
  \stackon[0pt]{#1}{\scalebox{.85}[1]{\color{red}$\sim$}}}

\usepackage{xcolor, soul}
\sethlcolor{yellow}

\begin{document}

\title{ A non-singular bouncing cosmology in $ f(R,T) $ gravity}

\author{J. K. Singh}
\email{jksingh@nsut.ac.in}
\affiliation{Department of Mathematics, Netaji Subhas University of Technology,\\ New Delhi-110 078, India}
\author{Shaily} 
\email{shaily.ma19@nsut.ac.in}
\affiliation{Department of Mathematics, Netaji Subhas University of Technology,\\ New Delhi-110 078, India}
\author{Akanksha Singh} 
\email{akanksha.ma19@nsut.ac.in}
\affiliation{Department of Mathematics, Netaji Subhas University of Technology,\\ New Delhi-110 078, India}
\author{Aroonkumar Beesham}
\email{abeesham@yahoo.com}
\affiliation{Department of Mathematical Sciences, University of Zululand, Kwa-Dlangezwa 3886, South Africa}
\affiliation{Faculty of Natural Sciences, Mangosuthu University of Technology, Umlazi 4031, South Africa}
\affiliation{National Institute for Theoretical and Computational Sciences (NITheCS), South Africa}
\author{Hamid Shabani}
\email{h.shabani@phys.usb.ac.ir}
\affiliation{Physics Department, Faculty of Sciences,  University of Sistan and Baluchestan, Zahedan, Iran}

\begin{abstract}
\qquad We investigate a bounce realization in the framework of higher order curvature in $ f(R,T) $ modified theory of gravity. We perform a detailed analysis of the cosmological parameters to explain the contraction phase, the bounce phase, and the expansion phase. Furthermore, we observe a violation of the null energy condition, instability of the model, and a singularity upon deceleration at the bouncing point, which are the supporting results for a bouncing cosmology. The outcome of the slow roll parameters is satisfactory to understand the inflation era and the equation of state parameter exhibits a ghost condensate behavior of the model near the bounce. Additionally, we discuss the stability of the model using linear perturbations in the Hubble parameter as well as the energy density.
\end{abstract}

\maketitle

\section{Introduction}\label{intro}
\qquad To study the different eras of the Universe, it is very important to analyze the underlying scenario from inflationary cosmology to structure formation. This is useful to observe the initial structure of the Universe. The present work in standard cosmology raises many issues related to inflationary cosmology like the initial singularity problem, and to avoid this problem, non-singular bounce cosmology is discussed by various authors. While the models related to bouncing cosmology face serious challenges like ghost and gradient instability problems \cite{Bars:2011mh, Brandenberger:2012zb, Cai:2012va, Cai:2014bea, Zhu:2021whu}, it is believed that the Universe can change its phase from contraction to expansion without encountering a singularity similar to the initial singularity in standard cosmology. As a result, the big bang cosmic singularity can be replaced by a  non-singular cosmic bounce, which shows a smooth transition from contraction to an expansion of the Universe. In a number of toy models, it is observed that the null energy condition (NEC) is violated in non-singular bouncing cosmology. Indeed, the violation of the NEC can occur in e.g., generalized Galileon theories that support the probability of non-singular cosmology. In recent times, big bounce cosmology has been a very interesting feature in the field of modified theories of gravity. In the literature, various kinds of bouncing cosmological models are discussed in different modified theories of gravity. The bounce cosmology in, inter alia, $ f(R) $ gravity \cite{Bamba:2014zoa, Saidov:2010wx, Barragan:2009sq, Amani:2015upn, Chakraborty:2018thg, Bamba:2013fha,Ilyas:2021xsu,Odintsov:2020zct}, $ f(R,T) $ gravity \cite{Agrawal:2021msm,Singh:2018xjv, Tripathy:2019nlw, Singh:2022jue,Bhardwaj:2022xjf}, $ f(R,G) $ gravity \cite{Sha, Nojiri:2022xdo, Elizalde:2020zcb, Barros:2019pvc, Terrucha:2019jpm}, $ f(T,B) $ gravity \cite{Caruana:2020szx}  has been  discussed. Also, smooth and slow phase transitions have also been studied by various authors. \\

In order to explain the bouncing scenario of the Universe, the following conditions will be examined. (1) The NEC should be violated as the Universe changes its phase from contraction to expansion ($ H $ should change sign),  \textit{i.e.}, in the neighborhood of the bouncing point, the NEC should be violated \cite{Singh:2018xjv,Cai:2007zv, Kobayashi:2016xpl}. (2) From the bouncing point ($ H=0 $), the scale factor starts increasing as the Universe enters its expansion phase. The deceleration parameter will be singular at $ H=0 $ as $ q=-1-\dot{H}/H^2 $. In late times, the deceleration parameter indicates cosmic acceleration. (3) In a successful bouncing model, the EoS parameter ($ w $) crosses the phantom line $ w=-1 $ near the transition point \cite{Cai:2007qw, Zhao:2006qg, Feng:2004ad}. For a successful bouncing model, these three conditions are required to be fulfilled. But, in some bouncing models like the Loop Quantum Bouncing model, the NEC needs not to be violated \cite{Elizalde:2019tee}. Other types of theories are called non-singular matter bounce scenarios which is a cosmological model with an initial state of matter-dominated contraction and a non-singular bounce \cite{Eas}. These models provide an alternative to inflationary cosmology for creating the observed spectrum of cosmological fluctuations \cite{Battefeld:2014uga, Cheung:2016oab, Brandenberger:2009jq, Wilson-Ewing:2012lmx}. In these theories, some matter fields are introduced in such a way that the WEC is violated in order to make $\dot{H}>0$ at the bounce. It is accessible that putting aside the correction term leads to negative values for the time derivative of the Hubble parameter for all fluids which respect WEC. Therefore, in order to obtain a bouncing cosmology it is necessary to either go beyond the GR framework or else to introduce new forms of matter which violate the key energy conditions, i.e. the null energy condition (NEC) and the WEC.\\
 
For a successful bounce, it can be shown that within the context of SCM the NEC and thus the WEC is violated for a period of time around the bouncing point. In the perspective of matter bounce scenarios, many authors have studied quintom matter \cite{Piao}, Lee–Wick matter \cite{Cai:2008qw}, ghost condensate field \cite{Bran}, Galileon field and phantom field \cite{Dzhunushaliev:2006xh, Nozari:2009zr, Saridakis:2010mf}. They have also constructed bouncing models using various approaches to modified gravity such as $f(R)$ gravity \cite{Oikonomou:2014jua, Odintsov:2014gea, Odintsov:2015zza}, teleparallel $f(T)$ gravity \cite{Cai:2011tc, Bamba:2016gbu}, Einstein–Cartan theory \cite{Brechet:2008zz, Poplawski:2010kb, Poplawski:2012ab, Poplawski:2011jz, Magueijo:2012ug, Hadi:2016zed} and others \cite{Sari}. Some other cosmological models such as Ekpyrotic model \cite{Khoury:2001wf} and string cosmology \cite{Gasperini:1992em, Florakis:2010is} which are alternatives to both inflation and matter bounce scenarios have also been discussed. To improve the consistency of the model with observations, a potential function is included in the action. For the flat FLRW space-time metric, a scalar field is associated with the potential function. From the background dynamics of the inflaton field, it is realized that during a sufficiently long inflationary phase, the slow roll parameters turn out to be much less than unity \cite{Cai:2010zma, Avelino:2012ue}.\\

 There are different motivations to include the trace of the energy-momentum tensor in the action; i) the existence of some exotic matter which affects the evolution of the Universe \cite{Shabani:2017rye, Shabani:2017kis} ii) there may be some unknown matter-gravity interactions iii) quantum effects in the form of conformal anomalies may work \cite{houndjo2013}. Cai \textit{et al.} \cite{Cai:2014bea} discuss a non-singular bounce cosmology with a single scalar field and matter. They explain the role of slow roll parameters as well as the spectral index and tensor-to-scalar ratio. Odintsov \textit{et al.} \cite{Elizalde:2014uba} also discusses many parameters to describe the inflaton cosmology in a bouncing scenario. Singh \textit{et al.} and Mishra \textit{et al} study the dynamical parameters of the bouncing Universe in $ f(R,T) $ gravity \cite{Singh:2018xjv, Agrawal:2021msm}. S. Nojiri \textit{et al.} \cite{Nojiri:2017ncd}, place an emphasis on bouncing cosmology in various modified theories of gravity. With the theory of bouncing cosmologies, an exact scale-invariant power spectrum of primordial curvature perturbations can be obtained along with a fine description. For example, during the contraction era, the matter bounce scenario is observed; during the expansion era, entropy is conserved and the perturbation modes increase with cosmic time. And this, the continuous cycle of cosmological bounces can be stopped if a crushing singularity takes place at the end of the expanding era. Odintsov \textit{et al.} studied a deformed matter bounce scenario \cite{Odintsov:2016tar}, where it was observed that the infinite repeating evolution of the Universe stops at the final attractor of the theory, which is a Big Rip singularity. \\

In this article, we discuss our work in the following order: Sec.~\ref{sectionEFE} starts with the action principle and basic formulas to discuss the Einstein field equations (EFE) for higher-order Ricci curvature. Further, to examine the bouncing scenario, we discuss some physical criteria, e.g., the cosmological parameters, the null energy condition, scalar field theory, and the stability of the model in Sec.~\ref{dynamical}. And afterward, in Sec.~\ref{conclusions}, we conclude our work with physically admissible results for a non-singular bouncing cosmology. In the following, we use units $ c = h= 1 $.
%

\section{Einstein Field Equations in $ f(R,T) $ gravity}\label{sectionEFE}
\qquad For modified $ f(R,T) $ gravity theory, the total gravitational action is given as
\begin{equation} \label{1}
\mathscr{A}=\int \Bigg[\frac{f(R,T)}{16 \pi G}+\mathscr{A}_m \Bigg]\sqrt{-g}  d^{4}x,
\end{equation} 
where $ \mathscr{A}_m $ is the matter Lagrangian, $ R $ is the  Ricci curvature, $ T $ is the trace of the energy momentum tensor (EMT), $ G $ is the gravitational constant and the function $ f(R,T) $  is taken to be the following coupling of $ R $ and $ T $, \textit{i.e.}, $ f(R,T)=f_1(R)+2f_2(T) $ \cite{Harko:2011kv}. By taking the variation in action (\ref{1}) \textit{w.r.t.} the metric tensor $ g_{ij} $, the gravitational equations leads to \cite{Singh:2022eun}
\begin{equation}\label{2}
f_1^R (R) R_{ij} - \frac{1}{2} g_{ij}(f_1(R)+2f_2(T))+(g_{ij} \Box -\nabla_i \nabla_j) f_1^R(R)=8\pi G T_{ij}-2f_2^T(T)(T_{ij}+\Theta_{ij}),
\end{equation}
where $ f_1^R (R) $ and $ f_2^T(T) $ represent the derivatives of functions $ f_1 $, $ f_2 $  \textit{w.r.t.} $ R $ and $ T $,  respectively. $ \nabla_i $ denotes the covariant derivative \textit{w.r.t} $ g_{ij} $, and $ \Box $ is the D'Alembert operator. $ \Theta_{ij} $ can be written as
\begin{equation}\label{3}
\Theta_{ij}=g^{\mu \nu} \frac{\delta{T_{\mu \nu}}}{\delta{g_{ij}}}=-2T_{ij}+g_{ij} \mathscr{A}_m-2g^{\mu \nu} \frac{\delta^2 \mathscr{A}_m}{\delta{g_{ij}} \delta{g_{\mu \nu}}}.
\end{equation}

Now, we assume, the matter Lagrangian $ \mathscr{A}_m =-p $ for a perfect fluid whose EMT is given as $ T_{ij} = (\rho+p)u_i u_j-p g_{ij} $, where $ \rho $ and $ p $ denote the energy density and pressure in the Universe. Using these relations, we obtain $ \Theta_{ij}=-2 T_{ij}-p g_{ij} $. Therefore, Eq. (\ref{2}) can be written as
\begin{equation}\label{4}
f_1^R (R) R_{ij} - \frac{1}{2} g_{ij}(f_1(R)+2f_2(T))+(g_{ij} \Box -\nabla_i \nabla_j) f_1^R(R)=8\pi G T_{ij}+2f_2^T(T)(T_{ij}+p g_{ij}).
\end{equation}
Eq. (\ref{4}) can be expressed in the standard form as
\begin{equation}\label{5}
G_{ij}=R_{ij}-\frac{1}{2} R g_{ij}= \frac{8 \pi G}{f_1^R(R)}(T_{ij}+T^*_{ij}).
\end{equation}
where \\
   $ ~~~~~~~~~~T^*_{ij}=\frac{1}{8 \pi G}\Big[\frac{1}{2} g_{ij}((f_1(R)+2f_2(T))-R f_1^R (R))+(\nabla_i \nabla_j-g_{ij} \Box) f_1^R(R)+2f_2^T(T)(T_{ij}+p g_{ij})\Big]$ \\\\
Next, we consider a homogeneous and isotropic flat Friedmann-Lemaitre-Robertson-Walker (FLRW) metric 
\begin{equation}\label{6}
ds^2=dt^2-a^2 (t)(dx^2+dy^2+dz^2).
\end{equation}
Here $ a(t) $ is the scale factor. From the EMT expression, the trace $ T $ can be calculated as
\begin{equation}\label{7}
T=\rho-3p,
\end{equation}
and the value of the Ricci scalar is
\begin{equation}\label{8}
R=-6(2 H^2+\dot{H}).
\end{equation}
By considering $ f_1(R)=R+\zeta R^m $ and $ f_2(T)=\lambda T^k $, we can find the Einstein field equations (EFE) as

\begin{equation}\label{1a}
3 H^2 = \frac{1}{1+\zeta mR^{m-1}} [ 8\pi\rho + \frac{1}{2} (\zeta(1-m)R^m+2\lambda T^k) - 3H\zeta m(m-1) R^{m-2} \dot{R} + (\rho +p)2\lambda k T^{k-1} ],	\nonumber
\end{equation}
or
\begin{eqnarray}
3 H^2 = \frac{1}{1+\zeta mR^{m-1}} [ 8\pi\rho + \frac{1}{2} (\zeta(1-m)(-6(2H^2+\dot{H}))^m+2\lambda T^k) ] + ~~~~~~~~~~~~~~~~~~~~~~~~~~   \nonumber   \\
\frac{1}{1+\zeta mR^{m-1}} [ - 3H\zeta m(m-1) (-6(2H^2+\dot{H}))^{m-2} (-6(4H\dot{H}+\ddot{H})) + (\rho +p)2\lambda k T^{k-1} ],	\nonumber
\end{eqnarray}

\begin{equation}\label{2a}
2\dot{H} + 3H^2 = \frac{-1}{a^2 (1+\zeta mR^{m-1})} [8\pi pa^2 + \frac{1}{2} (-a^2) (\zeta(1-m)R^m+2\lambda T^k ) + a\zeta m(m-1)R^{m-3} (a ((m-2)\dot{R}^2 + R\ddot{R}) + 3\dot{a}R\dot{R} ) ].	\nonumber
\end{equation}
or
\begin{eqnarray}
2\dot{H} + 3H^2 = \frac{-1}{a^2 (1+\zeta mR^{m-1})} [8\pi pa^2 + \frac{1}{2} (-a^2) (\zeta(1-m)(-6(2H^2+\dot{H}))^m+2\lambda T^k )+ ~~~~~~~~~~~~~~~~~~~~~~~~~~~~~~~~~~~~~~~~~~~~~~~~~~~~~~~~~~~~~~~~~~~~~~~~~ \notag \\   a\zeta m(m-1)(-6(2H^2+\dot{H}))^{m-3} (a ((m-2)(-6(4H\dot{H}+\ddot{H}))^2 + (-6(2H^2+\dot{H}))(-6(4H\ddot{H}+4\dot{H}^2+\dddot{H}))) + ~~~~~~~~~~~~~~~~~~~~~~~~~~~~~~~~~~~~\notag \\ 3\dot{a}(-6(2H^2+\dot{H}))(-6(4H\dot{H}+\ddot{H})) ) ]~~~~~~~~~~~~~~~~~~~~~~~~~~~~~~~~~~~~~~`.	\nonumber
\end{eqnarray}

Since the above equations will not be easy to solve thus for making the calculations easier, we have taken $ m=2 $ and $ k=1 $. Now, with $ f_1(R)=R+\zeta R^2 $, which was introduced by Starobinsky \cite{Starobinsky:1980te} and $ f_2(T)=\lambda T $, the Einstein field equations for the model will be 

\begin{equation}\label{9}
(8 \pi+3\lambda)\rho-\lambda p=3 H^2 +18 \zeta(\dot{H}^2-6 H^2 \dot{H}-2 H \ddot{H}),
\end{equation}
\begin{equation}\label{10}
(8 \pi+3 \lambda ) p-\lambda \rho=-2\dot{H}-3 H^2+6 \zeta(26 \dot{H} H^2+2\dddot{H}+14 H \ddot{H}+9\dot{H}^2).
\end{equation}

The energy density and pressure can be obtained from Eqs. (\ref{9}) and (\ref{10}) but for further discussion, a parametrization technique is needed. Thus, to draft a bouncing cosmological model with the conditions mentioned in Sec. \ref{intro}, we consider the bouncing scale factor as $ a(t) =(\alpha + \beta t^2)^{\frac{1}{n}} $ \cite{Navo:2020eqt}, where $ \alpha $, $ \beta $, and $ n $ are positive constants. And thereafter, using the relations $ H=\dot{a}/a $ and $ q=-1-\dot{H}/H^2 $, we acquire the value of the Hubble parameter and the deceleration parameter as follows:
\begin{align}
&H=\frac{2 \beta t}{n (\alpha+\beta t^2)}, \label{11}\\
&q=\frac{1}{2} \left(-\frac{\alpha  n}{\beta  t^2}+n-2\right).\label{12}
\end{align}

\begin{figure}

\begin{center}
\epsfig{figure=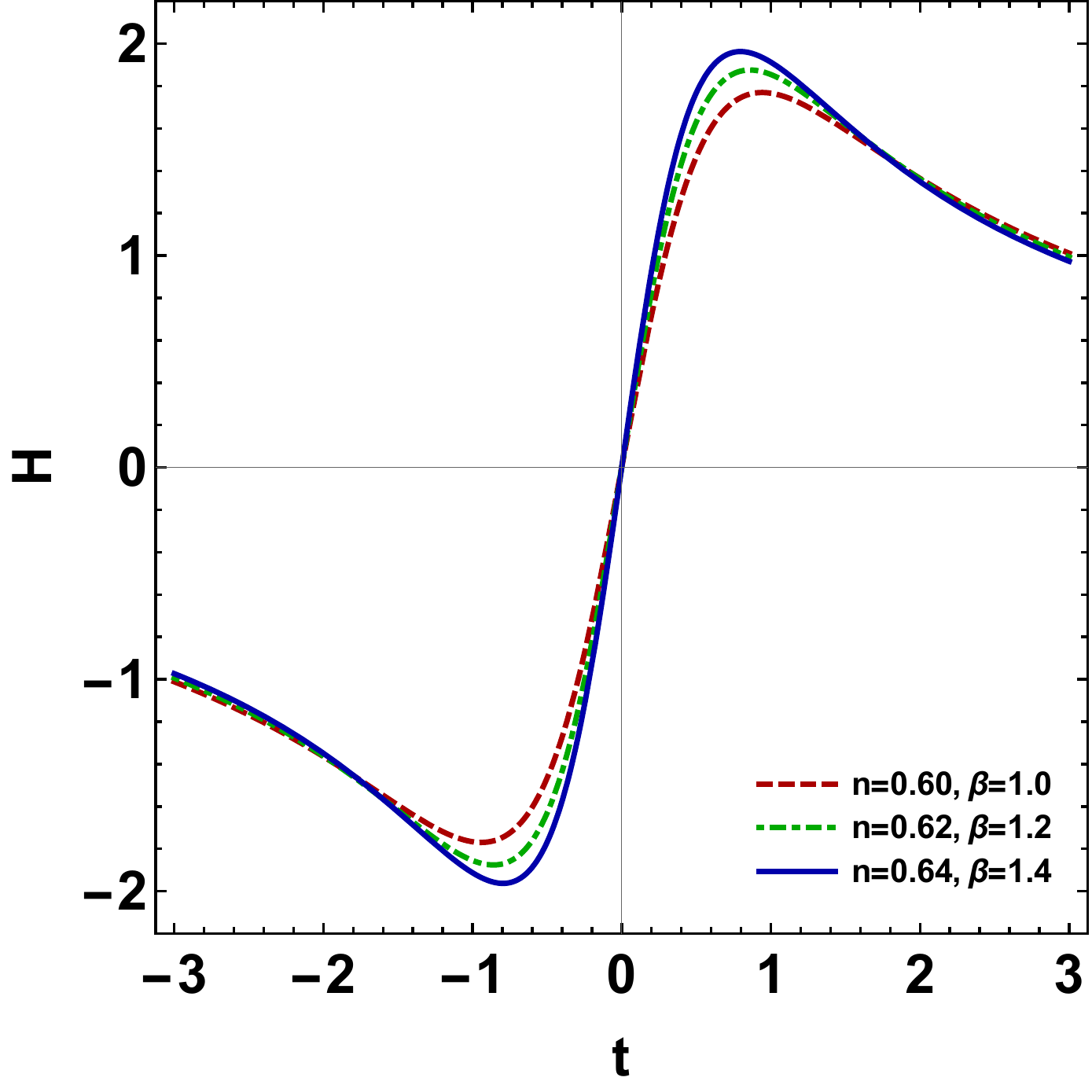,width=8.cm}\hspace{2mm}
\epsfig{figure=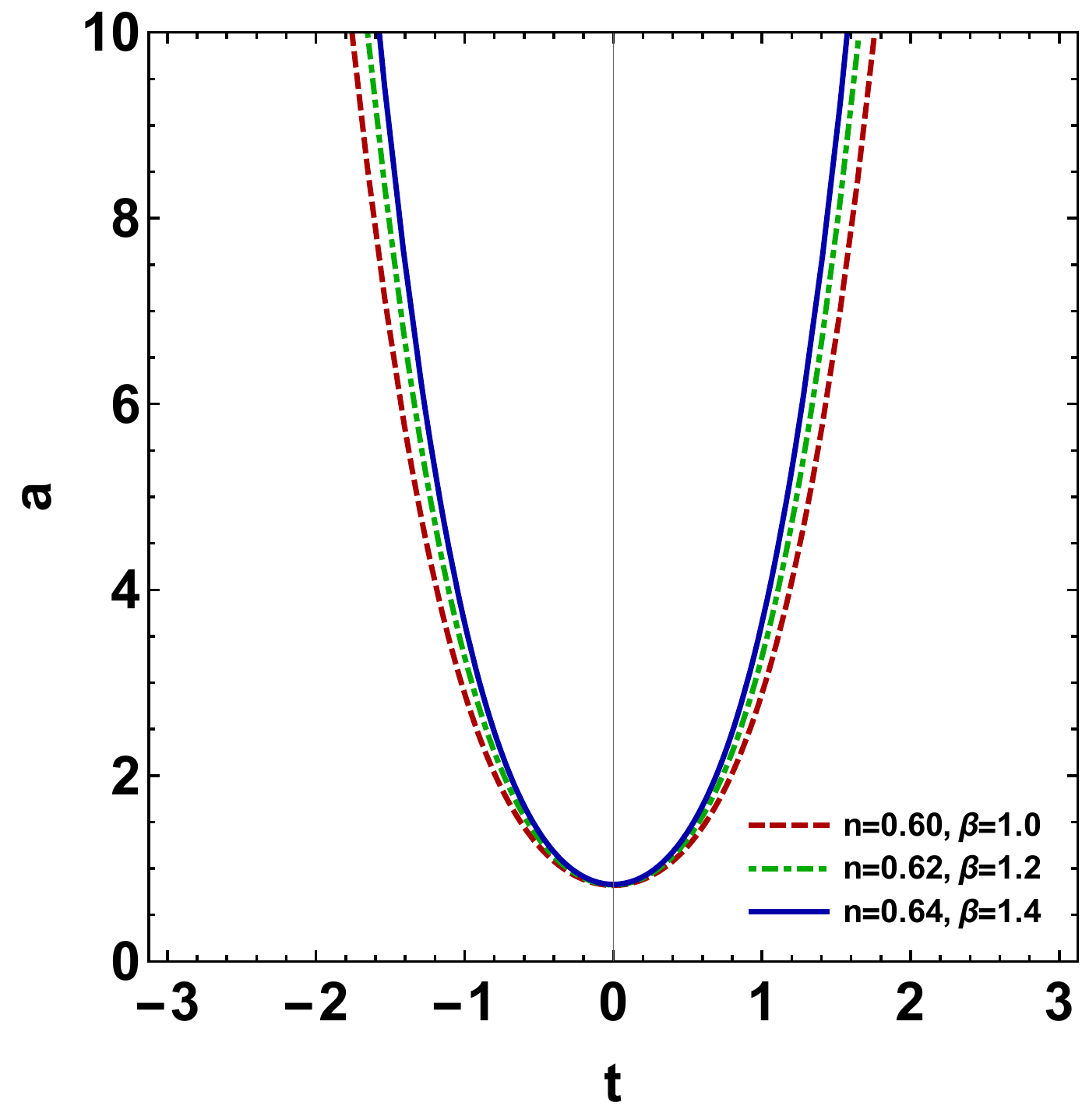,width=8.cm}\vspace{2mm}
\epsfig{figure=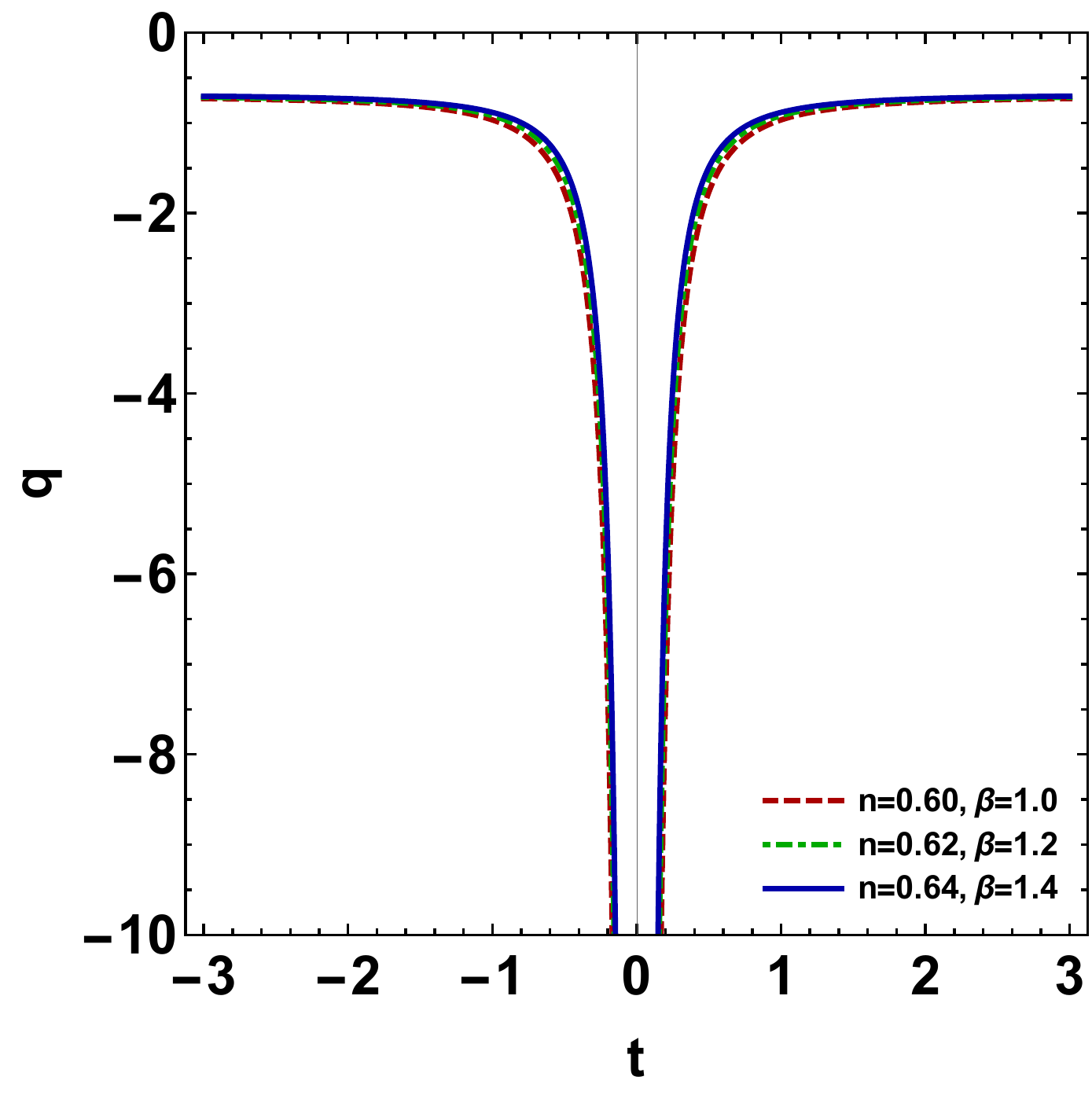,width=8.cm}
\caption{The graphical behavior of $ H $, $ a $ and $ q $. Note that the values $\alpha=0.887$, $\zeta=0.1$ and $\lambda=2.5$ have been used in Figs.~\ref{fig1}-\ref{fig6}. Also, we have used the same graphical options in these figures.}\label{fig1}
\end{center}
\end{figure}

From Eqs. (\ref{11}) and (\ref{12}) one finds that in the early and the late times the Hubble parameter vanishes and the deceleration parameter gets the limiting value $(n-2)/2$. Also, $H=0$, and $q$ diverges at the bouncing point. Since $\alpha$, $\beta$, and $n$ contain some positive values, to investigate the cosmological parameters we fix the value of $\alpha $ at $0.887$, and take small variations for the constants $\beta$ and $n$, which is mentioned in the figures. In Fig.~\ref{fig1}, all three plots of the Hubble parameter, scale factor, and deceleration parameter are favorable with the prescribed bouncing conditions. In the upper left panel of Fig.~\ref{fig1}, the Hubble parameter starts with a negative value (contracting phase), and afterward crosses the bouncing point ($H=0$) and enters the expanding phase ($H>0$). At the bouncing point, the scale factor has its least value and thereafter starts increasing (see the upper right panel of Fig.~\ref{fig1}). Also, the graphical behavior of the deceleration parameter can be noted in the lower panel of Fig.~\ref{fig1}. $q$ diverges at the singularity at the bouncing point and approaches the asymptotic value of $-1$ at late times. Furthermore, to analyze the dynamic behavior of the Universe, we are proceeding with the framed value of $a(t)$.   
%

\section{ Dynamical analysis of the bouncing evolution}\label{dynamical}
\quad The EoS parameter plays a pivotal role in the investigation of bounce cosmology. For this purpose, the energy density and pressure can be calculated from Eqs. (\ref{9}), (\ref{10}) as
\begin{align}\label{13}
\rho=\beta\frac{a_{1}t^{6}+a_{2}t^{4}+a_{3}t^{2}+a_{4}}{2 n^3(\lambda +2 \pi ) (\lambda +4 \pi )  (\alpha +\beta  t^2)^4},
\end{align} 
where
\begin{align}
&a_{1}=n\beta ^3 \big[\lambda  (n+6)+24 \pi \big],\nonumber\\
&a_{2}=\beta ^2 \Big\{48 \beta  \zeta  (7 \lambda +36 \pi )+\lambda  n^2 (\alpha -36 \beta  \zeta )+12 n \big[\lambda  (\alpha +5 \beta  \zeta )+4 \pi  (\alpha -9 \beta  \zeta )\big]\Big\},\nonumber\\
&a_{3}=\alpha  \beta \Big\{-48 \beta  \zeta  (7 \lambda +36 \pi )-\lambda  n^2 (\alpha -216 \beta  \zeta )+6 n \big[\lambda  (\alpha -12 \beta  \zeta )+4 \pi  (\alpha +60 \beta  \zeta )\big]\Big\},\nonumber\\
&a_{4}=n\alpha ^2 \Big\{36 \beta  \zeta  \big[4 \pi -\lambda  (n-3)\big]-\alpha  \lambda  n\Big\},\nonumber
\end{align} 
and
\begin{align}\label{14}
p=\beta \frac{b_{1}t^{6}+b_{2}t^{4}+b_{3}t^{2}+b_{4}}{2 \lambda  n^3(\lambda +2 \pi ) (\lambda +4 \pi )(\alpha +\beta  t^2)^4},
\end{align}
where
\begin{align}
&b_{1}=n\lambda\beta ^3 \big[3 \lambda  (n-2)+8 \pi  (n-3)\big],\nonumber\\
&b_{2}=\beta ^2 \Bigg\{3 \lambda ^2 (n-4) \big[\alpha (144 \beta +n)-12 \beta  \zeta  (3 n+7)\big]+8 \pi  \lambda  \big[324 \alpha  \beta  (n-4)+\alpha  (n-6) n-6 \beta  \zeta  (n-4) (6 n+41)\big],\nonumber\\
&+3456 \pi ^2 \beta  (n-4) (\alpha -\zeta )\Bigg\},\nonumber
\end{align}

\begin{align}
&b_{3}=\alpha  \beta  \Bigg\{-3 \lambda ^2\big[24 \beta  \zeta  \left(-9 n^2+3 n+14\right)+96 \alpha  \beta  (5 n-6)+\alpha  n (n+2)\big]-2304 \pi ^2 \beta  (5 n-6) (\alpha -\zeta ),\nonumber\\
&-8 \pi  \lambda\big[984 \beta  \zeta +216 \alpha  \beta  (5 n-6)+\alpha  n (n+3)-36 \beta  \zeta  n (6 n+13)\big]\Bigg\},\nonumber\\
&b_{4}=\alpha ^2 n \Bigg\{-1152 \pi ^2 \beta  (\alpha -\zeta )-3 \lambda ^2 \big[\alpha  (48 \beta +n)+36 \beta  \zeta  (n-3)\big]-8 \pi  \lambda  \big[\alpha  (108 \beta +n)+18 \beta  \zeta  (2 n-9)\big]\Bigg\}\nonumber,
\end{align}
and the EoS parameter ($w$) can be obtained from the equation of state as,
\begin{equation}\label{15}
w=\frac{p}{\rho}. 
\end{equation}

\begin{figure}
\begin{center}
\epsfig{figure=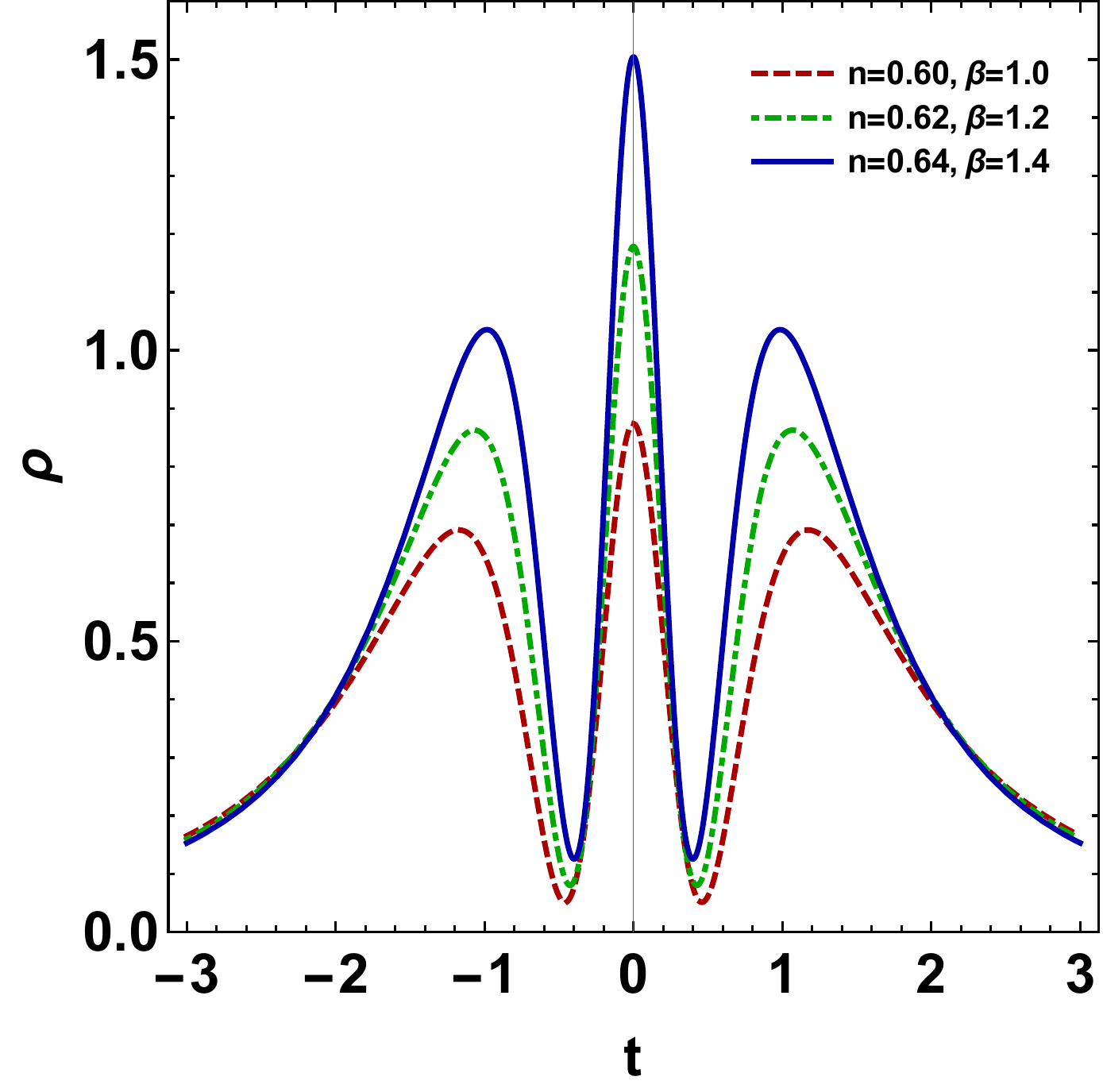,width=8.cm}
\epsfig{figure=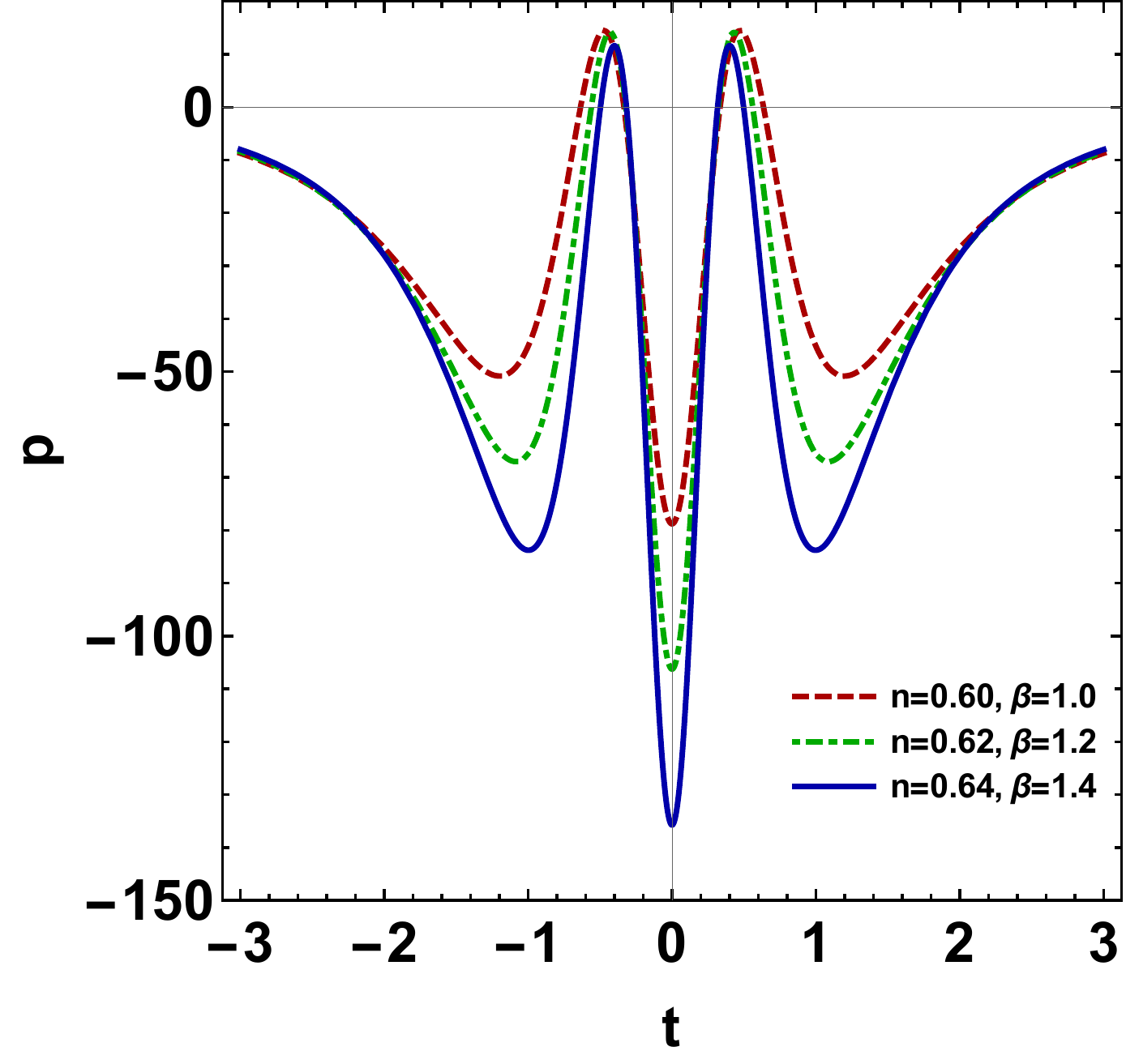,width=8.cm}
\epsfig{figure=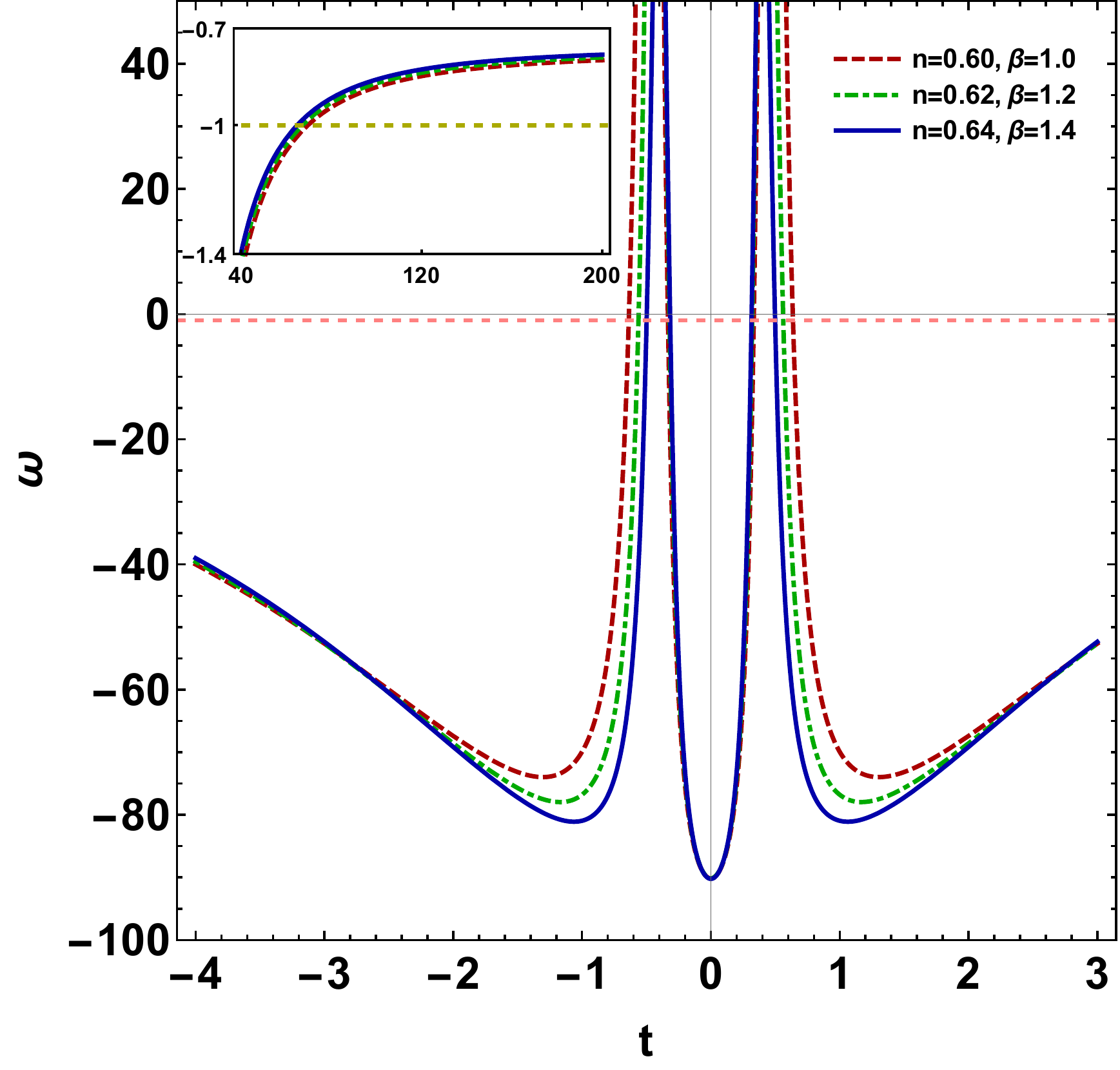,width=8.cm}
\caption{The plots of $\rho$, $p$ and $w$ \textit{w.r.t.} $t$.}\label{fig2}
\end{center}
\end{figure}

By considering Eqs. (\ref{13})-(\ref{15}) the following limiting results can be obtained
\begin{align}
&\lim_{t\to0} \rho=\beta\frac{ a_{4}}{d}, \mbox{\hspace{2cm}}    \lim_{t\to\infty} \rho=0,  \label{16}  \\
&\lim_{t\to0} p=\frac{\beta}{\lambda}\frac{ b_{4}}{ d}, \mbox{\hspace{2cm}}    \lim_{t\to\infty} p=0,   \label{17}   \\
&\lim_{t\to0} w=\frac{1}{\lambda}\frac{ b_{4}}{ a_{4}}, \mbox{\hspace{2cm}}    \lim_{t\to\infty} w=\frac{1}{\lambda}\frac{b_{1}}{ a_{1}},   \label{18}   \\
&d=2 \alpha ^4 (\lambda +2 \pi ) (\lambda +4 \pi ) n^3. \label{19}
\end{align}
Therefore, by fixing the model parameters $n$, $\alpha$, $\beta$, $\zeta$ and $\lambda$ one can model a bouncing scenario consistent with the theoretical requirements. In this regards, without loss of generality, the model parameters $\zeta$ and $\lambda$ are constrained to the values of 0.1 and 2.5, respectively. The graphical representation of the energy density ($\rho$), matter pressure ($p$) and the EoS parameter ($w$) can be seen in Fig.~\ref{fig2}. The energy density acquires its maximum value at the bouncing position, and just before and after it, it  drops (see the upper left of Fig.~\ref{fig2}). After a while, the value of $ \rho $ increases for a period and then decreases. In the late times, the  energy density is decreasing monotonically. In the upper right panel of Fig.~\ref{fig2}, the  matter pressure $p$ has its least negative value at the bouncing time. After a while, it starts increasing and becomes positive for a short interval of time. In the times, $p$ remains negative, which indicates the acceleration nature of the Universe. In the lower panel of Fig.~\ref{fig2}, it can be observed that the EoS parameter is very well shaped near the bouncing point. $w$ crosses the phantom line, $ w=-1 $, near the bounce in the late times, and again crosses the $\Lambda CDM$ line $(w=-1)$, and enters into a quintessence region, remaining there. At the bouncing point, our model is behaving like a ghost condensate model.   
%

\subsection{Violation of NEC}
\qquad Energy conditions come to light when one studies the Raychaudhuri equation for the expansion, which is completely a geometric statement and, in essence, it makes no commitment to any gravitational field equations \cite{Ashtekar:2011ni, Novello:2008ra, Giovannini:2017ndw}. Also, $ R_{ij}u^i u^j \geq 0 $ is obtained from the condition for attractive gravity. In general relativity (GR), we can write this condition in terms of the energy momentum tensor as $T_{ij} u^i u^j \geq 0$. However, in a non-singular bouncing model, a quintom context is associated with the violation of the null energy condition ($\rho+p \geq 0$), which is attained with various quantum instabilities \cite{Cai:2014bea}. 

\begin{figure}
\begin{center}
\epsfig{figure=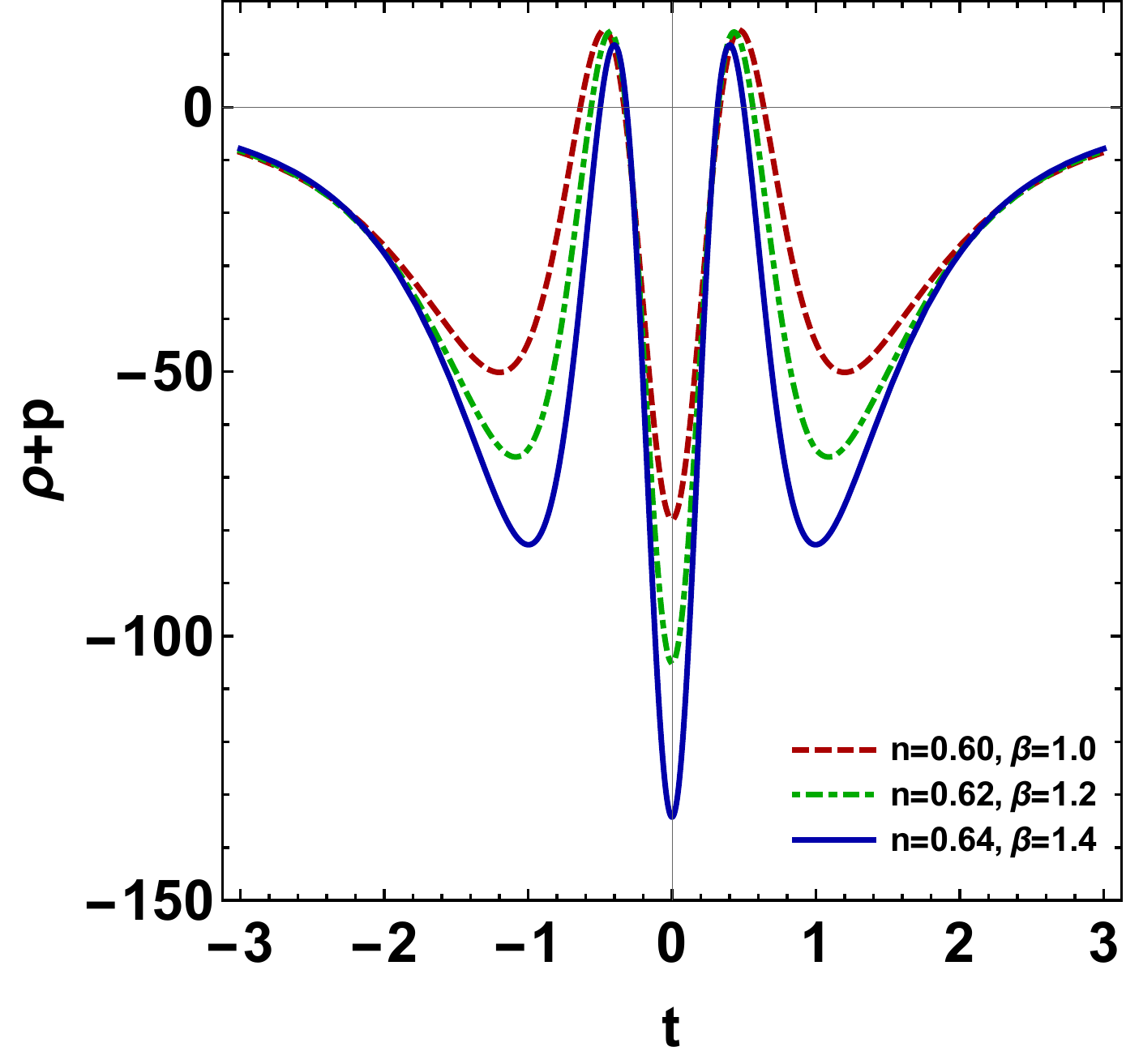,width=8cm}
\epsfig{figure=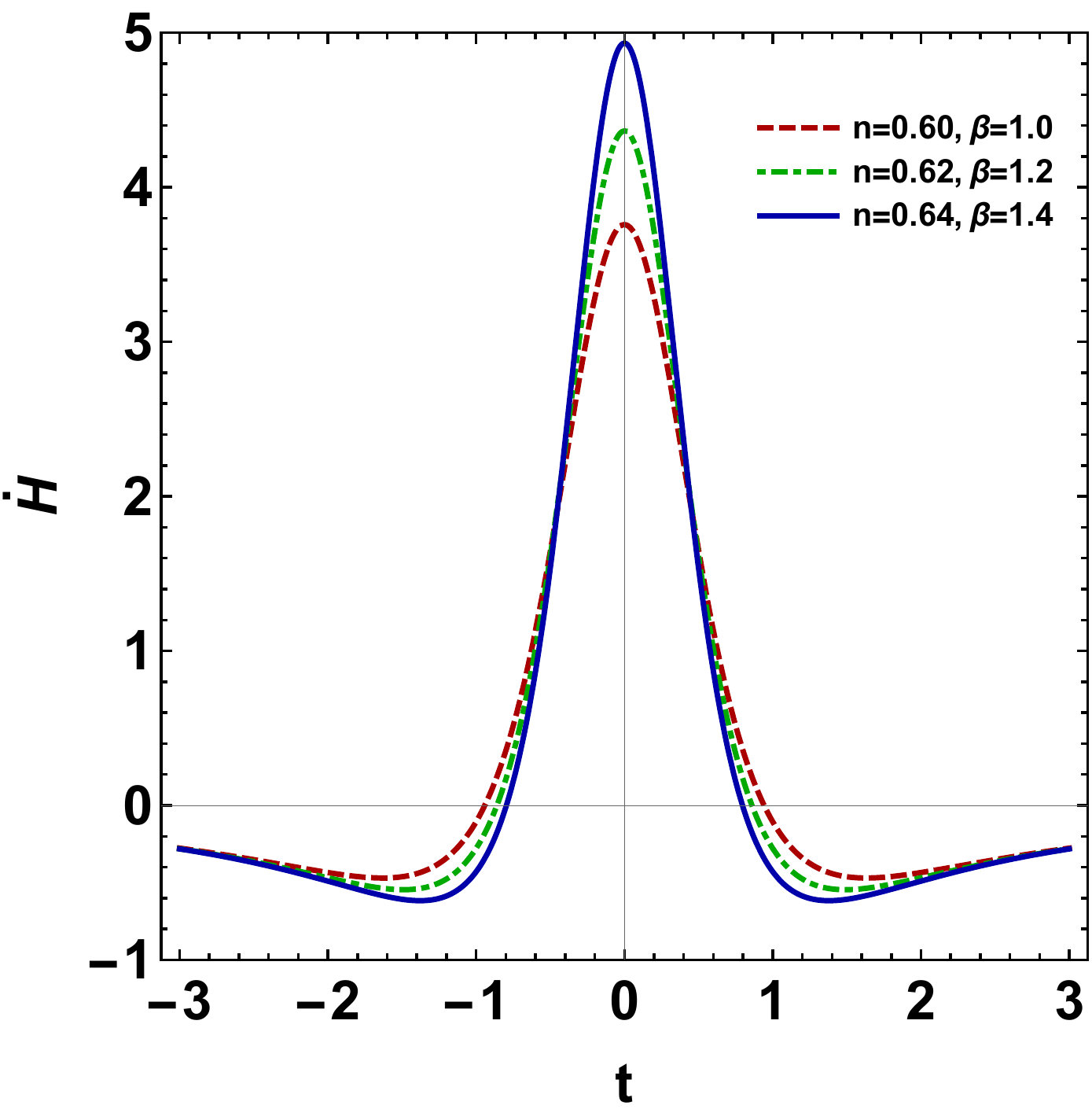,width=8cm}
\caption{The presentation of NEC as well as time variation of the Hubble parameter. One observes that $\dot{H}>0$ and $\rho+p<0$ in the neighbourhood of bouncing point. }\label{fig3}
\end{center}
\end{figure}

In our model, we can observe in the left panel of Fig.~\ref{fig3}, $ \rho+p <0 $, \textit{i.e.}, the NEC is  violated near the bounce, and in the right panel of Fig.~\ref{fig3}, we can observe that $\dot{H}$ is positive near the bounce, which supports the standard conditions \cite{Singh:2018xjv,Cai:2007zv, Kobayashi:2016xpl}.  
\subsection{Scalar field analysis}
\qquad In GR, it is interesting to study a bouncing cosmology in the quintom model. Generally, quintessence-like and phantom-like scalar fields are discussed in the quintom model. To make our model observationally consistent, it is required to take $w\approx -1$, which means the kinetic energy should be very much less as compared to the potential energy ($\dot{\phi}^2<<<V(\phi)$). 

To entice to reinterpret the source of matter as that of a scalar field, which is minimally coupled to gravity with a positive kinetic and potential term, and such kind of representation is also used by various authors \cite{Contreras:2017bqc, Barrow:1990vx, Chav}. Therefore, the present model can also be expressed as Friedmann universes containing a scalar field. Now, to discuss a non-singular bouncing cosmological model using scalar fields in $ f(R,T) $ gravity, let us consider a universe classified by the FLRW metric and a scalar field ($\phi$), with the action
\begin{equation}\label{20}
\mathscr{A}=\int \Bigg[\frac{R}{16 \pi G}-\frac{1}{2}\partial_{i}\phi\partial^{i}\phi-V(\phi)\Bigg]\sqrt{-g}  d^{4}x,
\end{equation}
where the scalar field $ \phi $ can be taken as quintessence-like or phantom-like. For the scalar field action, the EMT can be written as
\begin{equation}\label{21}
T_{ij}= \partial_i\phi \partial_j \phi-g_{ij}(\frac{1}{2}g^{ij} \partial_i\phi \partial_j \phi-V(\phi)),  
\end{equation}
and for the flat FLRW metric, we have $T_0^0=\rho $ and $ T_j^i=-p \delta_j^i$. Therefore the energy density and total pressure in the Universe filled by scalar fields (both quintessence-like and phantom-like), can be obtained as:
\begin{equation}\label{22}
\rho_{qu}=\frac{1}{2}\dot{\phi}_{qu}^2+V(\phi_{qu}), ~~~~~~ p_{qu}=\frac{1}{2}\dot{\phi}_{qu}^2-V(\phi_{qu})
\end{equation}
and
\begin{equation}\label{23}
\rho_{ph}=-\frac{1}{2}\dot{\phi_{ph}^2+V(\phi_{ph})}, ~~~~~~ p_{ph}=-\frac{1}{2}\dot{\phi_{ph}^2-V(\phi_{ph})},
\end{equation}
where the suffixes ``\textit{ph}" and ``\textit{qu}" denote phantom-like and quintessence-like scalar fields, respectively. 

From Eqs. (\ref{22}) and (\ref{23}), the kinetic and potential energies can be evaluated as:
\begin{align}
&\frac{1}{2}\dot{\phi}_{qu}^2=\rho_{qu}+p_{qu}, ~~~~~~ \frac{1}{2}\dot{\phi}_{ph}^2=-\rho_{ph}-p_{ph},\label{24}\\
&V(\phi_{qu})=\rho_{qu}-p_{qu}, ~~~~~~ V(\phi_{ph})=\rho_{ph}-p_{ph}.\label{25}
\end{align} 

\begin{figure}
\begin{center}
\epsfig{figure=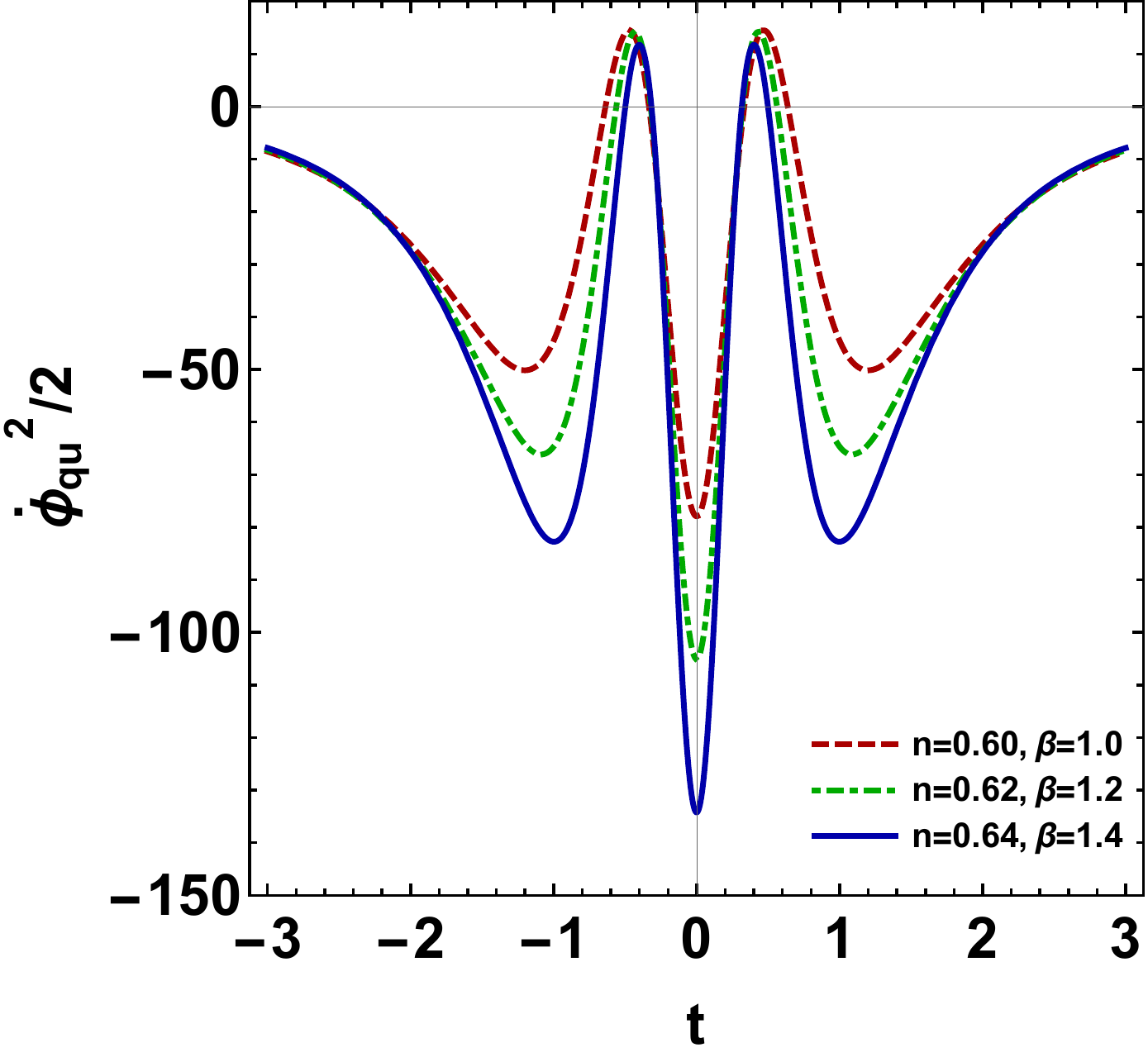,width=82mm}\hspace{2mm}
\epsfig{figure=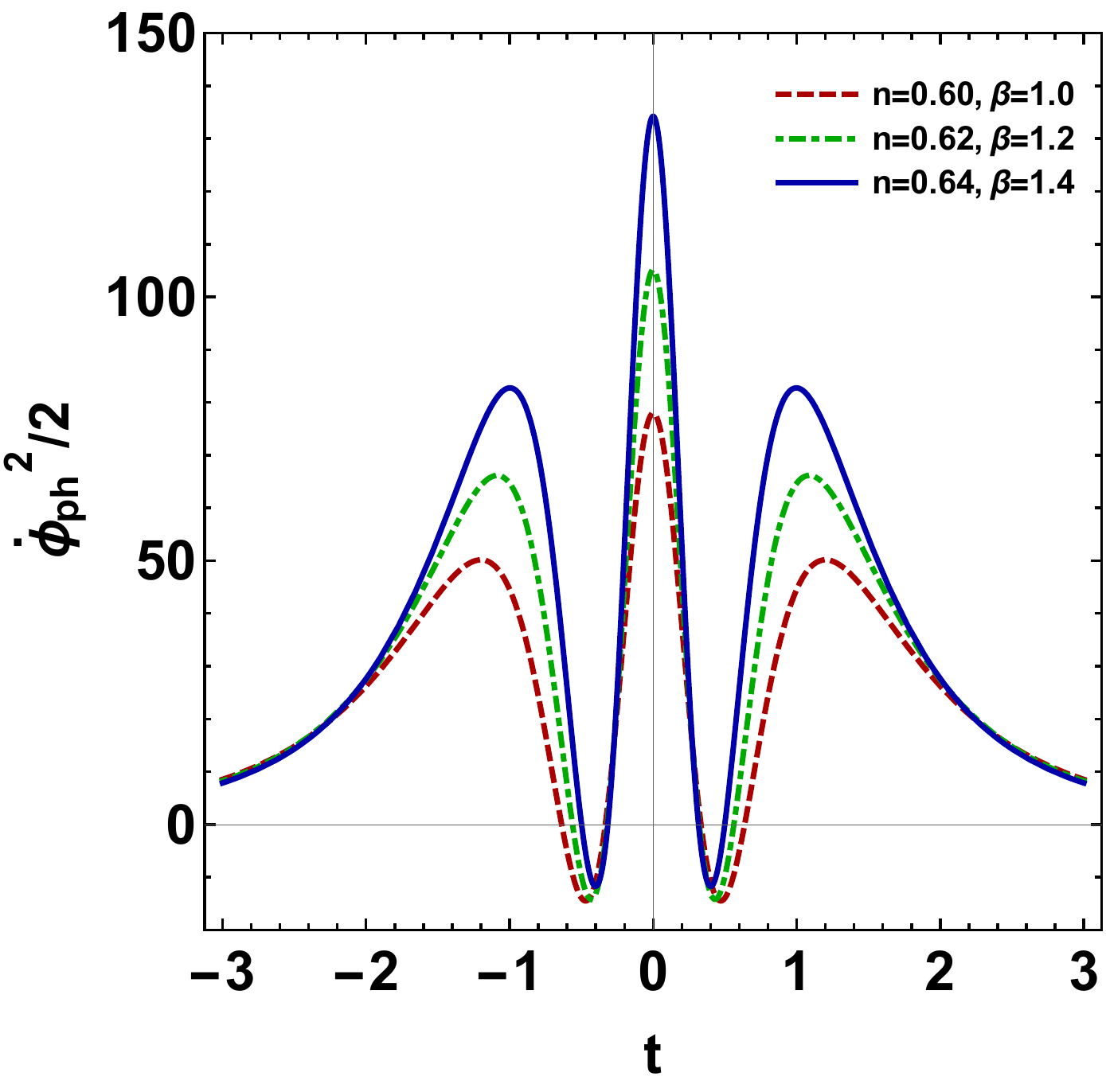,width=8cm}\vspace{2mm}
\epsfig{figure=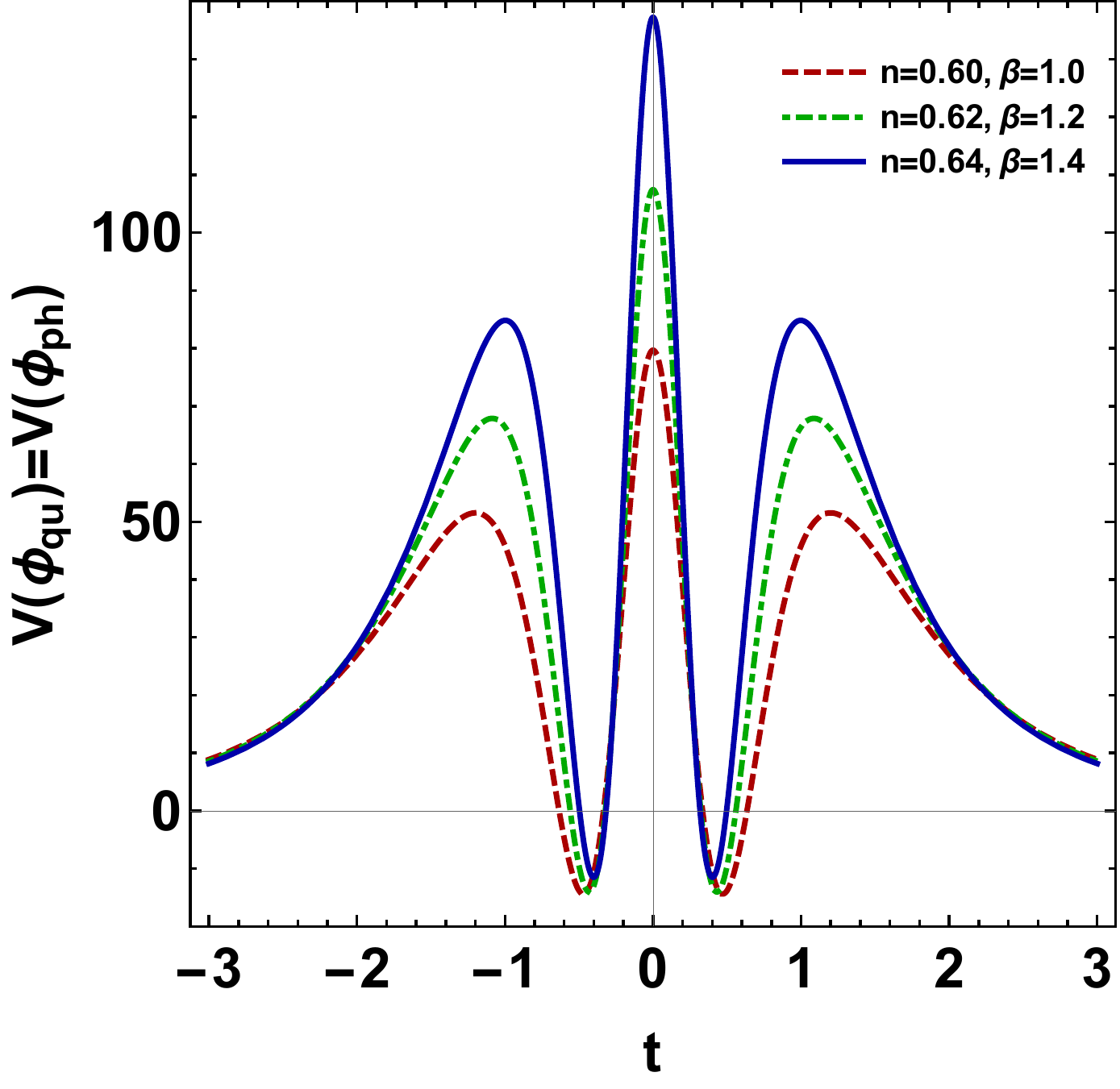,width=8cm}
\caption{The plots of kinetic and potential energies.}\label{fig4}
\end{center} 
\end{figure}

 Now, using the value of energy density and pressure from Eqs. (\ref{13}) and (\ref{14}), we can draw the plots for kinetic and potential energies (see Fig. ~\ref{fig4}).

In Fig.~\ref{fig4}, we depict the kinetic and potential energies, and analyse their variation for both quintessence-like and phantom-like scalar fields. Negative kinetic energy indicates the presence of dark energy. The upper left and right panels of Fig.~\ref{fig4} show that kinetic energy is negative for the quintessence-like scalar field, and positive for the phantom-like scalar field in the vicinity of the bouncing point. The potential energy for both scalar fields is equal and can be characterized from the lower panel of Fig.~\ref{fig4}. 

As is well-known, $w<-1$ for phantom-like scalar fields and $-1<w<0$ for quintessence-like scalar fields. Also, for the quintom  nature of the model, the kinetic energy for quintessence-like scalar fields ($\dot{\phi}_{qu}^2/2$) will be numerically equal to the kinetic energy for phantom-like scalar fields ($\dot{\phi}_{ph}^2/2$) when $ w $ crosses the quintom line $w=-1$, and this is the obligatory condition for having a bouncing model \cite{Shabani:2016dhj, Barrow:1988xh}. From the upper left and right panels of Fig.~\ref{fig4}, it can be observed that this condition holds good in the present model.\\ 

\begin{figure}
\begin{center}
\epsfig{figure=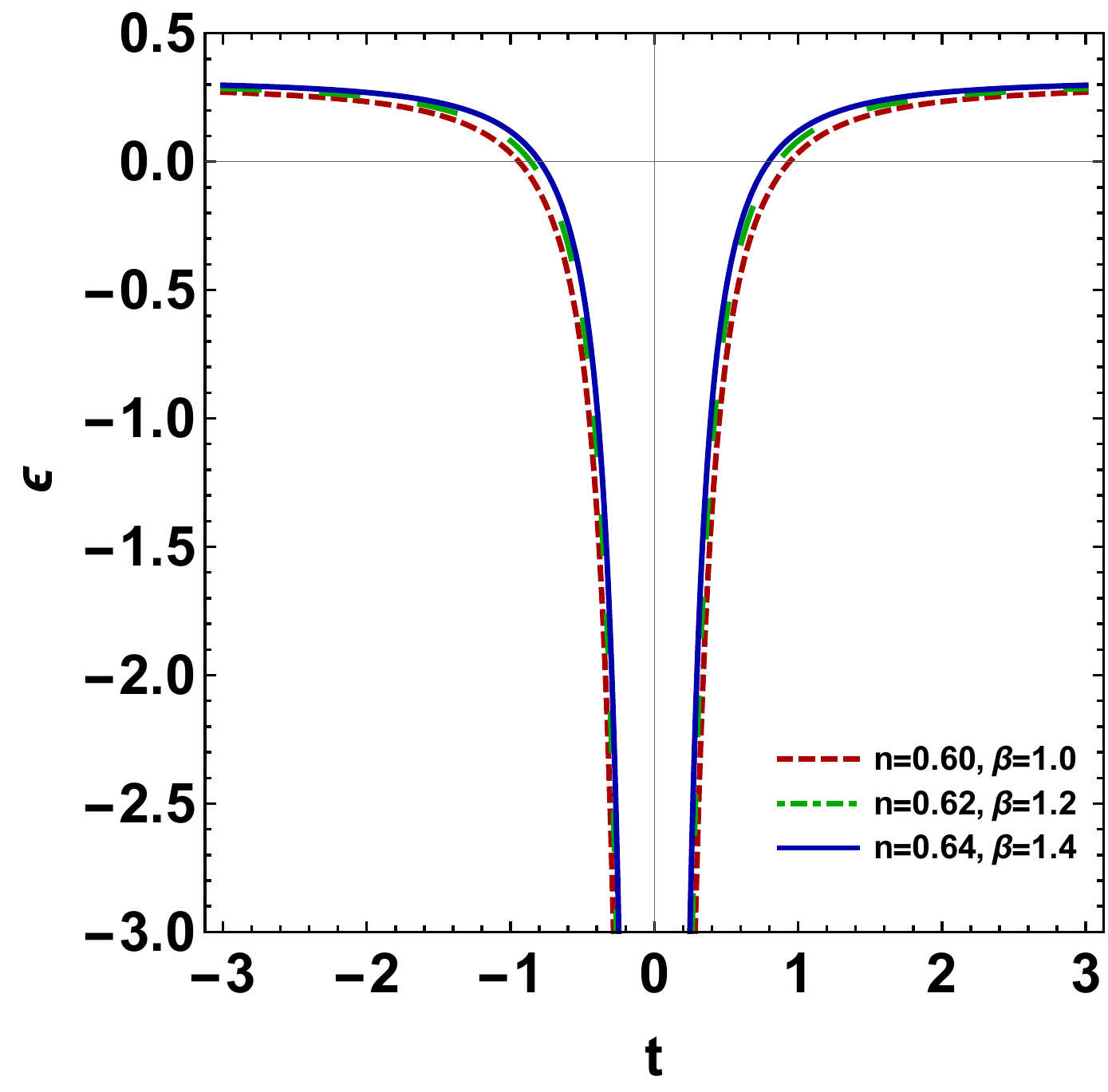,width=8cm}\hspace{2mm}
\epsfig{figure=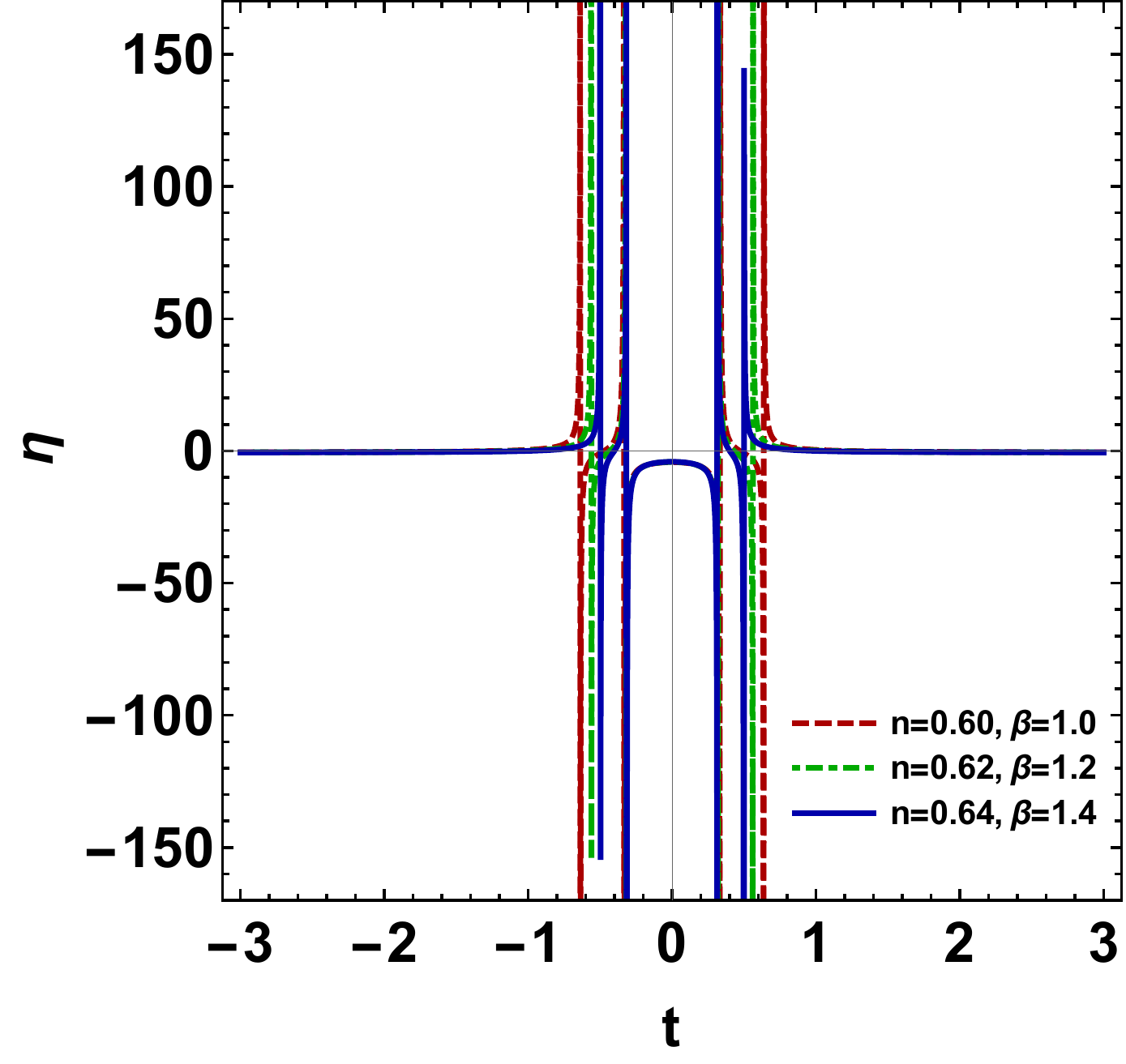,width=81mm}
\caption{The illustration of the behavior of the slow roll parameters.}\label{fig5}
\end{center}
\end{figure}

To study the inflation era, we can also analyse the behavior of the slow roll indices in scalar theories. Slow roll indices are defined as
\begin{equation}\label{26}
\epsilon=-\frac{\dot{H}}{H^2},		\mbox{\hspace{3em}}		\eta=\frac{\ddot{\phi}}{H \dot{\phi}}.
\end{equation}
Slow roll parameters for a scalar field should be much less than unity, which means that $\frac{1}{2}\dot{\phi}^2 << V(\phi)$.  Firstly, the slow-roll condition ensures the inflationary era in the first place, while the second ensures that inflation lasts for a sufficient amount of time \cite{Nojiri:2017ncd}. It is observed in Fig.~\ref{fig5} that  $\epsilon$ and $\eta$ obeys the required condition for the maximum time period.  
  
%

\subsection{Stability of the model}
\qquad We analyze the cosmological perturbation to study the evolution of the universe from the Big Bang. Several authors have worked on perturbation analysis for a homogeneous background cosmology \cite{Fry:1983cj,Bharadwaj:1996qm, delaCruz-Dombriz:2011oii, Bhardwaj:2022lrm, Jaybhaye:2022gxq, Narawade:2022jeg}. It is believed that due to a tiny perturbation in energy density, the cosmic fluid is not stable. Therefore, this concept gives a new sight into general relativity as it totally relates to the gravitational and pressure forces. Whenever the feeble pressure is equal to the gravitational force, then density perturbation is produced. Thus, to study the structure of the universe, we come across the analysis of small perturbations. In our work, we investigate linear perturbations to discuss the stability of the model. We inspect the linear perturbations in the Hubble parameter and energy density as
\begin{equation}\label{27}
    H_p(t) = H(t)(1+\delta(t))
\end{equation}
and
\begin{equation}\label{28}
    \rho_p(t) = \rho(t)(1+\delta_m(t)),
\end{equation}
where, $ H_p(t) $ and $ \rho_p(t) $ indicate the perturbed Hubble parameter and energy density respectively and $ \delta(t) $ and $ \delta_m(t) $ are taken as the perturbation terms of the Hubble parameter and the energy density respectively. Using the conservation equation for the matter field, the perturbation equation is obtained as
\begin{equation}\label{29}
    \dot{\delta_m}(t) + 3H(t)\delta(t) = 0.
\end{equation}
The expression that relates the geometrical and the matter perturbations is given as
\begin{equation}\label{30}
    b_m \delta_m(t) = -6\left[{H(t)}\right]^2\delta(t),
\end{equation}
where $ b_m = \kappa\rho_{m0} $. Matter perturbation measures the whole perturbation for a cosmological solution in GR. Now, by eliminating  $ \delta(t) $ from Eqs. (\ref{29}) and (\ref{30}), we obtain the perturbation equation of the first order as
\begin{equation}\label{31}
    \dot{\delta_m}(t) - \frac{b_m}{2H(t)}\delta_m(t) = 0.
\end{equation}
After integrating the Eq. (\ref{31}), we obtain
\begin{equation}\label{32}
    \delta_m(t) = D~exp\left[\frac{1}{2} \int \frac{b_m}{H(t)} dt\right].
\end{equation}
Similarly, the evolution of perturbation $ \delta(t) $ is given as
\begin{equation}\label{33}
    \delta(t) = -\frac{b_m}{6\left[{H(t)}\right]^2} D~exp\left[\frac{1}{2} \int \frac{b_m}{H(t)} dt\right].
\end{equation}
The terms $ \delta_m(t) $ and $ \delta(t) $ can be calculated as a function of redshift $z$ as
\begin{equation}\label{34}
    \delta_m(z) = D~exp\left[\frac{1}{2} \int \frac{b_m}{(1+z)~\left[{H(z)}\right]^2} dz\right]
\end{equation}
and
\begin{equation}\label{35}
    \delta(z) = -\frac{b_m}{6\left[{H(z)}\right]^2} D~exp\left[\frac{1}{2} \int \frac{b_m}{(1+z)~\left[{H(z)}\right]^2} dz\right].
\end{equation}
where $ D $ is an arbitrary integration constant.

\begin{figure}\centering
	\subfloat[]{\label{a}\includegraphics[scale=0.41]{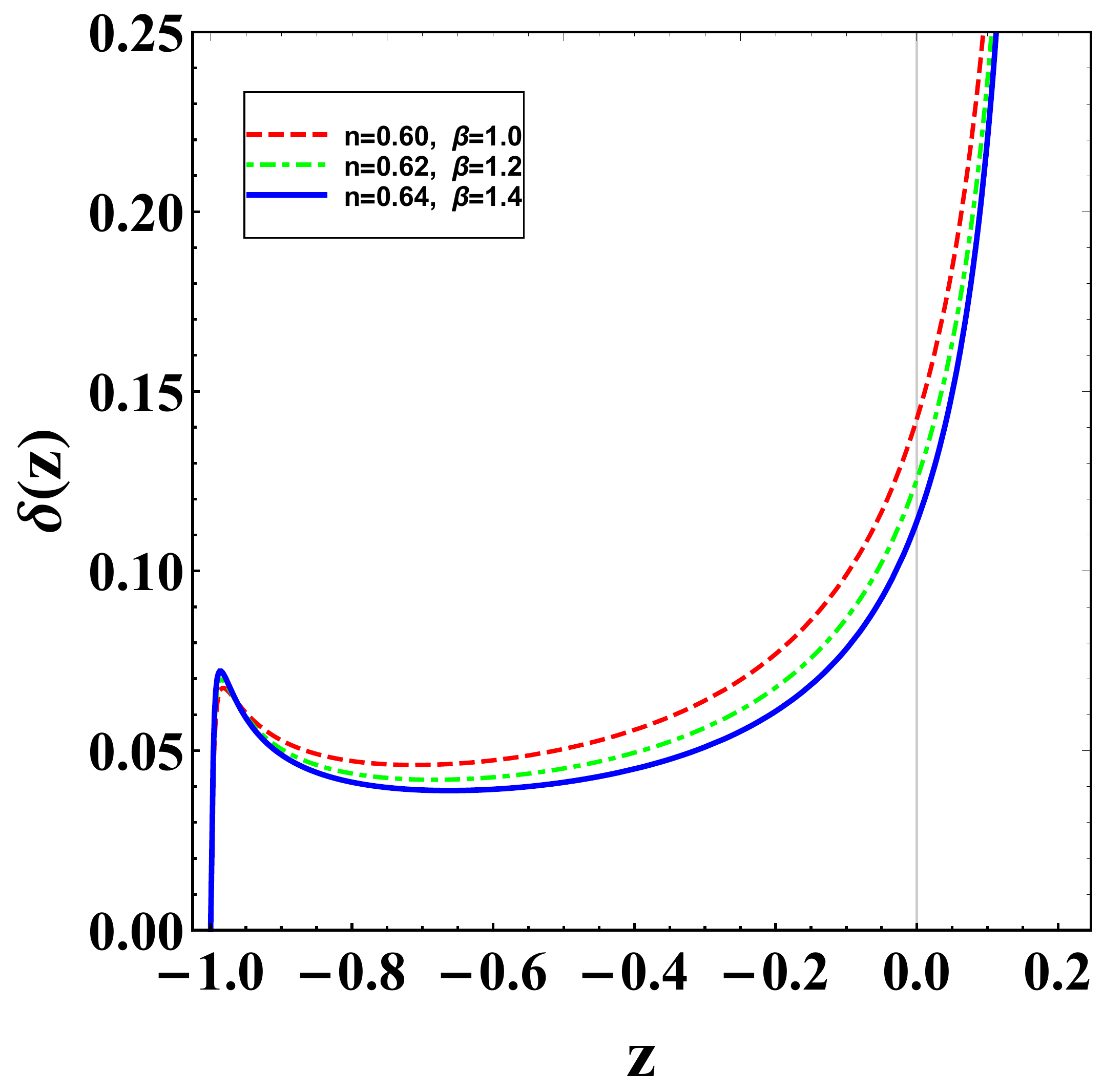}}\hfill
	\subfloat[]{\label{b}\includegraphics[scale=0.4]{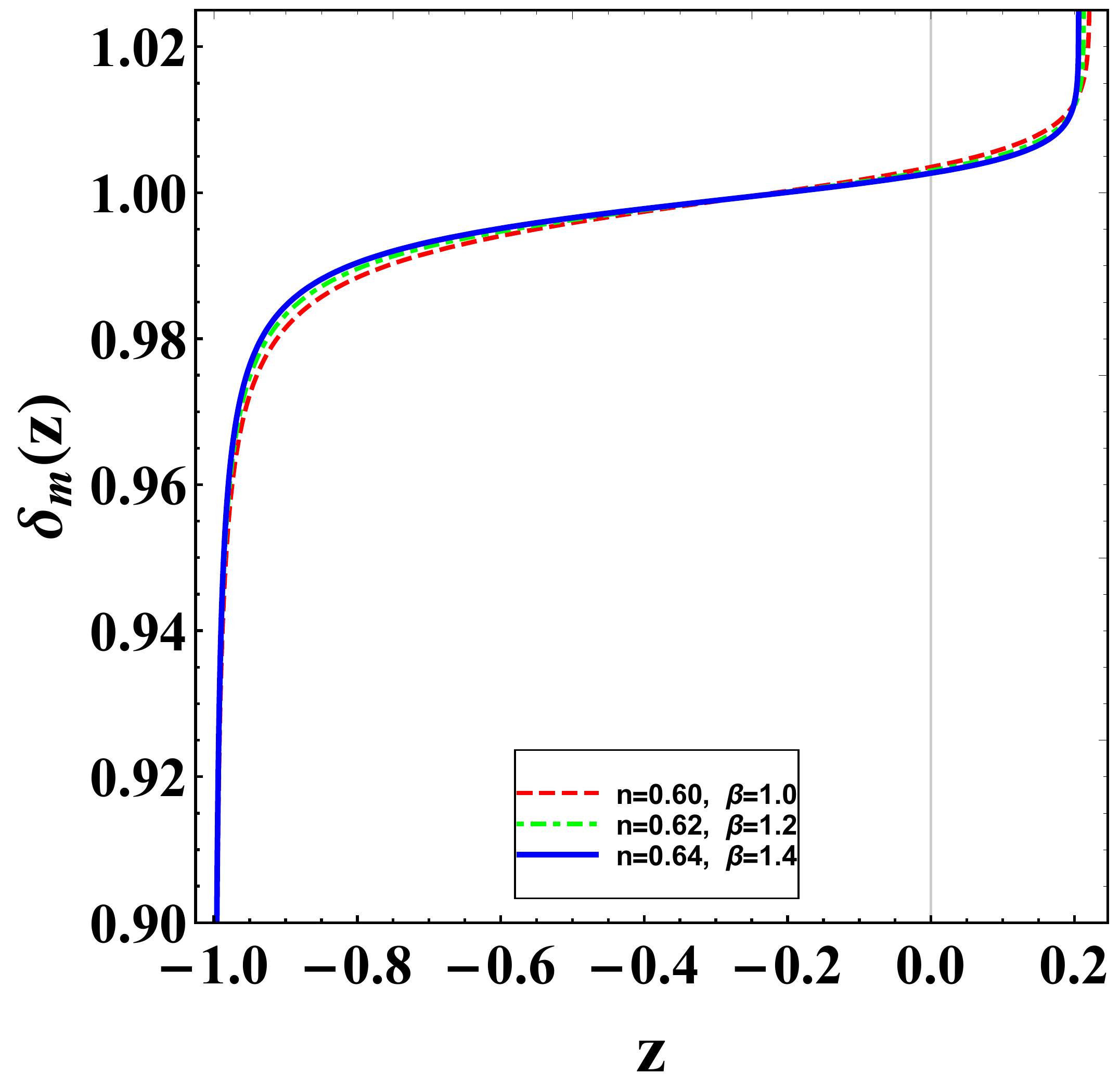}} 
	\caption{\scriptsize Plots for perturbation $ \delta(z) $ in Hubble parameter and $ \delta_m(z) $ in energy density vs. redshift $ z $} \label{fig6}
\end{figure}

In Fig. \ref{fig6}, we see that the perturbation parameters $ \delta(z) $ and $ \delta_m(z) $ decrease monotonically and tend to zero as $ z\to -1 $. Therefore, our model seems to be stable under linear perturbation in the late times. 
%

\section{Conclusion}\label{conclusions}
\quad To scrutinise the theoretical inference of non-singular bouncing cosmologies, it is required to break a series of singularity theorems manifested by many authors. Among them, one is a violation of the NEC in the framework of GR. In the literature, a non-singular bouncing cosmology in the early Universe could be related with the quintom scenario as stimulated by the dark energy study of the late time acceleration \cite{Cai:2009zp}. 

In this paper, we have presented a model in modified gravity for a flat FLRW space-time by the parametrization of the scale factor, which yields a non-singular bouncing cosmology and if we take $\lambda=0$ or $\zeta=0$, the bouncing scenario does not change. From the functioning of the dynamical parameters, our model is experiencing a cosmic bounce at $ t=0 $. At this point, the Hubble parameter indicates the phase from contraction to expansion. Also, the required conditions for a successful bouncing model are examined. The proposed model violates the NEC and has a singularity at the bouncing point as shown by the deceleration parameter. The behavior of the EoS parameter also supports the bouncing cosmological model as $w$ crosses the phantom line $w=-1$ near $t=0$, and shows a ghost condensate behavior near the bounce. In the early times and the late times, the EoS parameter crosses the quintom line again.       

In a scalar field theory, we observe that the kinetic energies are equal at the time when $w$ crosses the quintom line, which is also a supportive result to have a bounce. The slow roll parameters show satisfactory output in this proposed model. The calculated values of the tensor-to-scalar ratio and the spectral index are studied in $f(R,T)$ theory of gravity and both values are not near the recently improved limit on the tensor-to-scalar ratio presented by the Planck team. For a number of e-folds, $ N = 60 $, the value of $\delta_m=1-n_s=2(3\epsilon-\eta)$ is positive and very close to zero and the value of $r=16\epsilon$ are negative in our model while r is a positive quantity. This shows that the Planck data alone falsify the polynomial inflationary models independently of the observational restriction on r \cite{Gron:2022lqj}. Finally, we examine the stability of the model via linear perturbation, and we notice that our model is stable (see Fig.~\ref{fig6}). Thus, we conclude that the model shows a non-singular bounce and is stable in the late times.   
\vskip0.2in 
\textbf{\noindent Acknowledgements} Shaily and Akanksha Singh express their thanks to Prof. J. P. Saini, Hon’ble Vice Chancellor, NSUT, New Delhi for the fellowship under TRFs scheme. 

\vskip0.2in
\textbf{\noindent Data Availability Statement} No new data were created or analysed in this study.

\end{document}